\documentclass[twocolumn]{aastex631}

\usepackage{xspace}
\usepackage{CJK}

\hypersetup{
    colorlinks=true,
    linkcolor=blue,
    filecolor=magenta,      
    urlcolor=cyan,
}

\usepackage[normalem]{ulem} 
\usepackage{cancel} 

\usepackage{color}

\newcommand{\um}{\ensuremath{\mu\rm{m}}\xspace}
\newcommand{\kms}{\ensuremath{\rm{km\,s}^{-1}}\xspace}

\newcommand{\siggas}{\ensuremath{\Sigma_{\rm{gas}}}\xspace}
\newcommand{\uJy}{\ensuremath{\mu\rm{Jy}}\xspace}

\newcommand{\lir}{\ensuremath{L_{\rm{IR}}}\xspace}

\newcommand{\Mdust}{\ensuremath{M_{\rm{dust}}}\xspace}

\newcommand{\Mgas}{\ensuremath{M_{\rm{gas}}}\xspace}

\newcommand{\Mdyn}{\ensuremath{M_{\rm{dyn}}}\xspace}
\newcommand{\Mhalo}{\ensuremath{M_{\rm{halo}}}\xspace}
\newcommand{\Msol}{\ensuremath{\rm{M}_\odot}\xspace}

\newcommand{\cii}{[C{\scriptsize II}]\xspace}

\newcommand{\arc}{\ensuremath{''}\xspace}

\shortauthors{J.~S.~Spilker, et~al.}
\shorttitle{Clumpy Massive Galaxy Formation at $z\sim7$}

\begin{document}
\begin{CJK*}{UTF8}{gbsn}

\defcitealias{marrone18}{M18}

\title{Chaotic and Clumpy Galaxy Formation in an Extremely Massive Reionization-Era Halo}

\correspondingauthor{Justin S. Spilker}
\email{jspilker@tamu.edu}

\author[0000-0003-3256-5615]{Justin~S.~Spilker}
\affiliation{Department of Physics and Astronomy and George P. and Cynthia Woods Mitchell Institute for Fundamental Physics and Astronomy, Texas A\&M University, 4242 TAMU, College Station, TX 77843-4242, US}

\author[0000-0003-4073-3236]{Christopher~C.~Hayward}
\affiliation{Center for Computational Astrophysics, Flatiron Institute, 162 Fifth Avenue, New York, NY, 10010, USA}

\author[0000-0002-2367-1080]{Daniel~P.~Marrone}
\affiliation{Steward Observatory, University of Arizona, 933 North Cherry Avenue, Tucson, AZ 85721, USA}

\author[0000-0002-6290-3198]{Manuel~Aravena}
\affiliation{N\'{u}cleo de Astronom{\'i}a de la Facultad de Ingenier\'{i}a y Ciencias, Universidad Diego Portales, Av. Ej\'{e}rcito Libertador 441, Santiago, Chile}

\author[0000-0002-3915-2015]{Matthieu~B{\'e}thermin}
\affiliation{Aix Marseille Univ., CNRS, CNES, LAM, Marseille, France}

\author[0000-0001-8706-1268]{James~Burgoyne}
\affiliation{Department of Physics and Astronomy, University of British Columbia, 6225 Agricultural Rd., Vancouver, V6T 1Z1, Canada}

\author{Scott~C.~Chapman}
\affiliation{Department of Physics and Astronomy, University of British Columbia, 6225 Agricultural Rd., Vancouver, V6T 1Z1, Canada}
\affiliation{National Research Council, Herzberg Astronomy and Astrophysics, 5071 West Saanich Rd., Victoria, V9E 2E7, Canada}
\affiliation{Department of Physics and Atmospheric Science, Dalhousie University, Halifax, Nova Scotia, Canada}

\author[0000-0002-2554-1837]{Thomas~R.~Greve}
\affiliation{Cosmic Dawn Center, DTU Space, Technical University of Denmark, Elektrovej 327, Kgs. Lyngby, DK-2800, Denmark}
\affiliation{Department of Physics and Astronomy, University College London, Gower Street, London WC1E6BT, UK}

\author{Gayathri~Gururajan}
\affiliation{Aix Marseille Univ., CNRS, CNES, LAM, Marseille, France}

\author[0000-0002-8669-5733]{Yashar~D.~Hezaveh}
\affiliation{D\'{e}partement de Physique, Universit\'{e} de Montr\'{e}al, Montreal, Quebec, H3T 1J4, Canada}
\affiliation{Center for Computational Astrophysics, Flatiron Institute, 162 Fifth Avenue, New York, NY, 10010, USA}

\author{Ryley~Hill}
\affiliation{Department of Physics and Astronomy, University of British Columbia, 6225 Agricultural Rd., Vancouver, V6T 1Z1, Canada}

\author[0000-0002-4208-3532]{Katrina~C.~Litke}
\affiliation{Steward Observatory, University of Arizona, 933 North Cherry Avenue, Tucson, AZ 85721, USA}

\author[0000-0001-7964-5933]{Christopher~C.~Lovell}
\affiliation{Centre for Astrophysical Research, School of Physics, Astronomy \& Mathematics, University of Hertfordshire, Hatfield AL10 9AB, UK}

\author[0000-0001-6919-1237]{Matthew~A.~Malkan}
\affiliation{Department of Physics and Astronomy, University of California, Los Angeles, CA 90095-1547, USA}

\author[0000-0001-7089-7325]{Eric~J.~Murphy}
\affiliation{National Radio Astronomy Observatory, 520 Edgemont Rd, Charlottesville, VA 22903, USA}

\author[0000-0002-7064-4309]{Desika~Narayanan}
\affiliation{Department of Astronomy, University of Florida, 211 Bryant Space Sciences Center, Gainesville, FL 32611, USA}
\affiliation{University of Florida Informatics Institute, 432 Newell Drive, CISE Bldg E251, Gainesville, FL 32611, USA}
\affiliation{Cosmic Dawn Center, DTU Space, Technical University of Denmark, Elektrovej 327, Kgs. Lyngby, DK-2800, Denmark}

\author[0000-0001-7946-557X]{Kedar~A.~Phadke}
\affiliation{Department of Astronomy, University of Illinois, 1002 West Green St., Urbana, IL 61801, USA}

\author[0000-0001-7477-1586]{Cassie~Reuter}
\affiliation{Department of Astronomy, University of Illinois, 1002 West Green St., Urbana, IL 61801, USA}

\author[0000-0002-2718-9996]{Antony~A.~Stark}
\affiliation{Center for Astrophysics $\vert$ Harvard \& Smithsonian, 60 Garden Street, Cambridge, MA 02138, USA}

\author[0000-0002-3187-1648]{Nikolaus~Sulzenauer}
\affiliation{Max-Planck-Institut f\"{u}r Radioastronomie, Auf dem H\"{u}gel 69, D-53121 Bonn, Germany}

\author[0000-0001-7192-3871]{Joaquin~D.~Vieira}
\affiliation{Department of Astronomy, University of Illinois, 1002 West Green St., Urbana, IL 61801, USA}
\affiliation{Department of Physics, University of Illinois, 1110 West Green St., Urbana, IL 61801, USA}
\affiliation{National Center for Supercomputing Applications, University of Illinois, 1205 West Clark St., Urbana, IL 61801, USA}

\author[0000-0001-7610-5544]{David~Vizgan}
\affiliation{Department of Astronomy, University of Illinois, 1002 West Green St., Urbana, IL 61801, USA}
\affiliation{Cosmic Dawn Center, DTU Space, Technical University of Denmark, Elektrovej 327, Kgs. Lyngby, DK-2800, Denmark}

\author[0000-0003-4678-3939]{Axel~Wei{\ss}}
\affiliation{Max-Planck-Institut f\"{u}r Radioastronomie, Auf dem H\"{u}gel 69, D-53121 Bonn, Germany}

\begin{abstract}

The SPT0311-58 system at $z=6.900$ is an extremely massive structure within the reionization epoch, and offers a chance to understand the formation of galaxies in an extreme peak in the primordial density field. 
We present 70\,mas Atacama Large Millimeter/submillimeter Array observations of the dust continuum and \cii 158\,\um emission in the central pair of galaxies and reach physical resolution $\sim$100-350\,pc, among the most detailed views of any reionization-era system to date. 
The observations resolve the source into at least a dozen kiloparsec-size clumps. 
The global kinematics and high turbulent velocity dispersion within the galaxies present a striking contrast to recent claims of dynamically cold thin-disk kinematics in some dusty galaxies just 800\,Myr later at $z\sim4$.
We speculate that both gravitational interactions and fragmentation from massive parent disks have likely played a role in the overall dynamics and formation of clumps in the system.
Each clump individually is comparable in mass to other $6 < z < 8$ galaxies identified in rest-UV/optical deep field surveys, but with star formation rates elevated by $\sim3-5\times$. 
Internally, the clumps themselves bear close resemblance to greatly scaled-up versions of virialized cloud-scale structures identified in low-redshift galaxies.
Our observations are qualitatively similar to the chaotic and clumpy assembly within massive halos seen in simulations of high-redshift galaxies.

\end{abstract}

\section{Introduction} \label{intro}

The vast majority of known $z>6$ galaxies have been found through optical color selection techniques that identify predominantly low-mass, UV-bright star-forming galaxies, but the rarity and strong clustering of more massive galaxies typically exclude them from all but the widest survey fields. Far-infrared surveys offer an alternative perspective on the early universe, identifying the UV-dim population of dusty star-forming galaxies (DSFGs). Of these only a handful have been identified into the reionization epoch at $z>6$, likely inhabiting dark matter halos that are much more rare and massive than UV-selected galaxies \citep[e.g.][]{cooray14,strandet17,zavala18}. With star formation rates (SFRs) of hundreds of \Msol/yr, these IR-luminous dusty galaxies most likely then become the first massive quiescent galaxies now spectroscopically confirmed out to $z\sim4$ \citep[e.g.][]{glazebrook17,straatman16,forrest20b}. While the DSFG population has long been argued to consist primarily of major mergers \citep[e.g.][among many others]{ivison98,tacconi08}, recent works have found other $z\sim4$ DSFGs that appear to have kinematics dominated by thin-disk rotation \citep[e.g.][]{rizzo21,fraternali21}, possible progenitors of dynamically cold, rotating quiescent galaxies at high redshift \citep[e.g.][]{toft17,newman18}.

The $z=6.900$ DSFG SPT0311-58 is currently both the highest-redshift IR-selected system and a very massive object within the epoch of reionization (\citealt{strandet17}; \citealt{marrone18}, hereafter \citetalias{marrone18}; \citealt{jarugula21}). Resolved into a pair of massive dusty galaxies by $\sim$0.3'' ALMA observations, the masses of dust ($\log \Mdust/\Msol \approx 9-9.5$) and cold gas ($\log \Mgas/\Msol \approx 11-11.5$) are extreme compared to other known reionization-era galaxies. Even under conservative assumptions and considering only the currently-known spectroscopically-confirmed members, the implied host dark matter halo mass is $\log \Mhalo/\Msol \approx 12-13$ just 800\,Myr after the Big Bang, among the most massive halos expected in the standard $\Lambda$CDM cosmology over an area of tens of square degrees \citepalias{marrone18}. This structure offers the chance to probe the details of galaxy assembly within an extreme peak of the primordial density field.

Here we present new 0.07'' resolution ALMA observations of the dust and \cii\,158\,\um emission in SPT0311-58, improving on the spatial resolution of the previous observations from \citetalias{marrone18} by $\sim$16$\times$ in beam area. The data reveal a highly clumpy and turbulent structure in the rapidly star-forming central galaxies, among the most detailed views of any reionization-era system. The observations and analysis are described in Section~\ref{data}, and our main findings are discussed in Section~\ref{results}. We conclude and describe future lines of work in Section~\ref{conclusions}. We assume a flat $\Lambda$CDM cosmology with $\Omega_m=0.307$ and $H_0=67.7$\,\kms\,Mpc$^{-1}$ \citep{planck15}; for this cosmology, 1'' $=$ 5.4\,kpc at $z=6.900$. Images and data products are available at \url{https://github.com/spt-smg/publicdata}.

\section{Data and Methods} \label{data}

\begin{figure*}
\begin{centering}
\includegraphics[width=\textwidth]{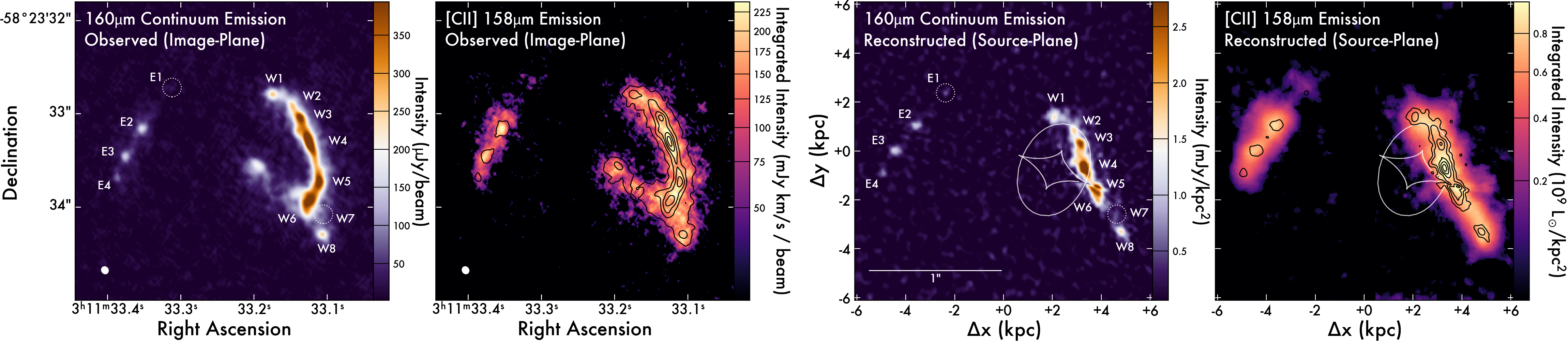}
\end{centering}
\caption{
Observed (image-plane) and reconstructed (source-plane) maps of the dust and integrated \cii emission in SPT0311-58. The 12 clumps we identify in the source-plane reconstructions are labeled; clumps E1 and W7 are identified only in the \cii data cube and are marked with dotted circles. Because of the relatively simple lensing geometry, each of the source-plane clumps map nearly one-to-one with the clumps visible by eye in the image plane (clumps W2--W4 together create the southeastern arc/counterimage). Contours in the \cii maps are of the dust continuum. The $\approx$0.07\arc synthesized beam is shown in the lower-left of the observed images, and the lensing caustics are shown in the source-plane maps in white.
}\label{fig:data}
\end{figure*}

\subsection{ALMA Observations}

ALMA observations of the dust and \cii emission were carried out in projects 2016.1.01293.S and 2017.1.01423.S, designed to reach spatial resolutions $\approx$0.25\arc and 0.05\arc, respectively, at 240\,GHz. The 2016 data were presented in \citetalias{marrone18}, and the new data are essentially identical in spectral setup. The spectral resolution of the sideband containing the \cii line is $\approx$10\,\kms. The newly-acquired data were obtained between 2017 November and 2018 October, comprising 40\,min on-source in a compact array configuration with 15\,m--1.4\,km baseline lengths and 65\,min in an extended array with 0.1--8.5\,km baselines. The data were reduced using the standard ALMA pipeline and combined with the previous 2016 data. We performed a single round of phase-only self-calibration with a solution interval equal to the scan length. We imaged both the continuum and \cii data using Briggs weighting with robustness parameter 0.5; the resulting synthesized beam size is 0.07\arc$\times$0.08\arc. The continuum image reaches 7.5\,\uJy/beam sensitivity, and the \cii data cube typically reaches 85\,\uJy/beam sensitivity in 40\,\kms channels. The continuum image is shown in Fig.~\ref{fig:data}, together with an integrated \cii (moment-0) map made by integrating channels from $-600$ to $+1120$\,\kms with respect to $z=6.900$. As noted by \citetalias{marrone18}, SPT0311-58 consists of a western (SPT0311-58W) and eastern (SPT0311-58E) component, with SPT0311-58W mildly lensed by a foreground galaxy and SPT0311-58E essentially unlensed. The two components are separated by $\approx$700\,\kms and are therefore distinct objects \citepalias[][and see below]{marrone18} that our new data resolve in detail.

\subsection{Lensing Reconstruction}

Lens models and source-plane reconstructions of $\approx$0.3\arc ALMA data, including the \cii line, were originally presented in \citetalias{marrone18}. Those models used a pixellated reconstruction of the source plane and represented the foreground lens as a singular isothermal ellipsoid (SIE), following the methodology described in \citet{hezaveh16}. We use the same code and reconstruction procedure for our high-resolution data. We re-fit for the foreground lens parameters including all available data, finding good quantitive agreement with the previous models. We reconstructed the continuum emission and \cii emission in 40\,\kms channels, applying the best-fit lensing deflections. The peak residuals from the best-fit reconstruction reach $\sim$4$\sigma$ in the continuum, compared to a peak S/N$\approx$120 in the data. This demonstrates that our assumption of an SIE profile, while simple, is adequate in the absence of other data to constrain the lens mass profile. We expect these residuals to have minimal impact on our subsequent analysis. The reconstruction of the continuum and integrated \cii emission are shown in Fig.~\ref{fig:data}, while Fig.~\ref{fig:ciicube} shows a more detailed 3D rendering of the \cii emission. Because SPT0311-58E lies far from the multiply-imaged region of the source plane and is essentially unlensed, its apparent morphology is reproduced basically exactly in the reconstruction. Similarly, much of the apparent image-plane structure in SPT0311-58W maps simply to the intrinsic source plane, a consequence of the spatial offset between lens and source. Due to variations in the local magnification, the effective spatial resolution of the reconstructions varies from $\approx$60\,pc (near the lensing caustics in the central regions of SPT0311-58W) to $\approx$350\,pc (across SPT0311-58E and the southern end of SPT0311-58W, which are not strongly lensed).

\subsection{Clump Identification}\label{clumpid}

The source-plane reconstructions of continuum and \cii emission both show a highly clumpy structure. To verify this structure and measure clump properties, namely the size and spectral line width of each clump, we use the \texttt{fellwalker} clump-finding algorithm \citep{berry15}, which identifies distinct peaks in two- or three-dimensional data and integrates outward from these peaks to a user-specified minimum threshold. We analyze the continuum and \cii data separately, i.e. we do not impose any correspondence between continuum and \cii clumps. The \texttt{fellwalker} algorithm does not presume any spatial or spectral profile for the clumps. It instead simply labels the flux of each pixel as belonging to at most a single clump; any spatial or spectral overlap between clumps is therefore ignored. We require clumps to reach a minimum peak signal-to-noise of 4 in the continuum and 8 in the \cii cube, and expand around peaks identified this way down to the 2$\sigma$ level. We also tested several other clump-finding techniques, including the classic \texttt{clumpfind} algorithm \citep{williams94} and the tree-based \texttt{astrodendro} package\footnote{\url{http://www.dendrograms.org/}}. All methods yielded consistent results for the continuum clumps; the clump-to-clump contrast in the continuum is high enough that the details of the clump identification are not very important. There was slightly more variation in the results from the \cii cube due to the lower contrast between adjacent clumps. The most common differences were a failure to identify the clump we label E1 due to its faintness, and the joining of clumps W3 and W4 due to the low contrast between clumps. With some fine-tuning of the parameters we were able to recover all clumps consistently between the different algorithms, but we stress that our results are not strongly dependent on the details of the clump identification.

We identify 12 distinct clumps in the \cii data cube, four in SPT0311-58E (labeled E1--E4 from north to south) and eight in SPT0311-58W (W1--W8). Ten clumps are also identified in the continuum image. Clump E1 peaks just below our minimum continuum S/N threshold, while clump W7 appears to be genuinely continuum-faint. For these two clumps we `manually' measure continuum properties using apertures over the \cii-emitting region; we verified that this method would also accurately recover the continuum flux density of the other clumps as well. All continuum clumps spatially overlap with a \cii counterpart, and we conclude that the continuum clumps are physically associated with the corresponding \cii clumps. Together the continuum (\cii) clumps recover $\approx$80\% ($\approx$70\%) of the total intrinsic emission measured by integrating within a large aperture over the source plane. The remainder is distributed in a handful of additional tentative clumps that do not reach our minimum peak signal-to-noise threshold, and a diffuse component surrounding the algorithmically-identified clumps that falls below our 2$\sigma$ threshold and is therefore not assigned to any particular clump. The \cii spectra of all clumps, the sum of clumps, and the sum over the entire source plane are shown in Figure~\ref{fig:ciispectra}.

We tested the recovery of clump properties by injecting artificial Gaussian clumps into signal-free regions of the reconstructed data cube and then measuring the clump properties in the same way as the real data. For artificial clumps with properties similar to the real ones (radius R$\sim$1\,kpc, peak line flux $\sim$2.5\,mJy), the size and integrated line fluxes were recovered well. The line width was also well-recovered as long as FWHM$\gtrsim$40\,\kms (the cube channel width, not coincidentally). This implies the minimum line width recoverable in our data is $\sigma \approx 17$\,\kms, which places limits on our later comparisons to clumps detected at low redshifts with narrow line widths.


\section{Results and Discussion} \label{results}

\subsection{Clumpy Formation of a Massive z\,$\sim$\,7 Galaxy} \label{kinematics}

\begin{figure*}
\begin{centering}
\begin{interactive}{js}{spt0311_ciicube_interactive.tar.gz}
\includegraphics[width=0.9\textwidth]{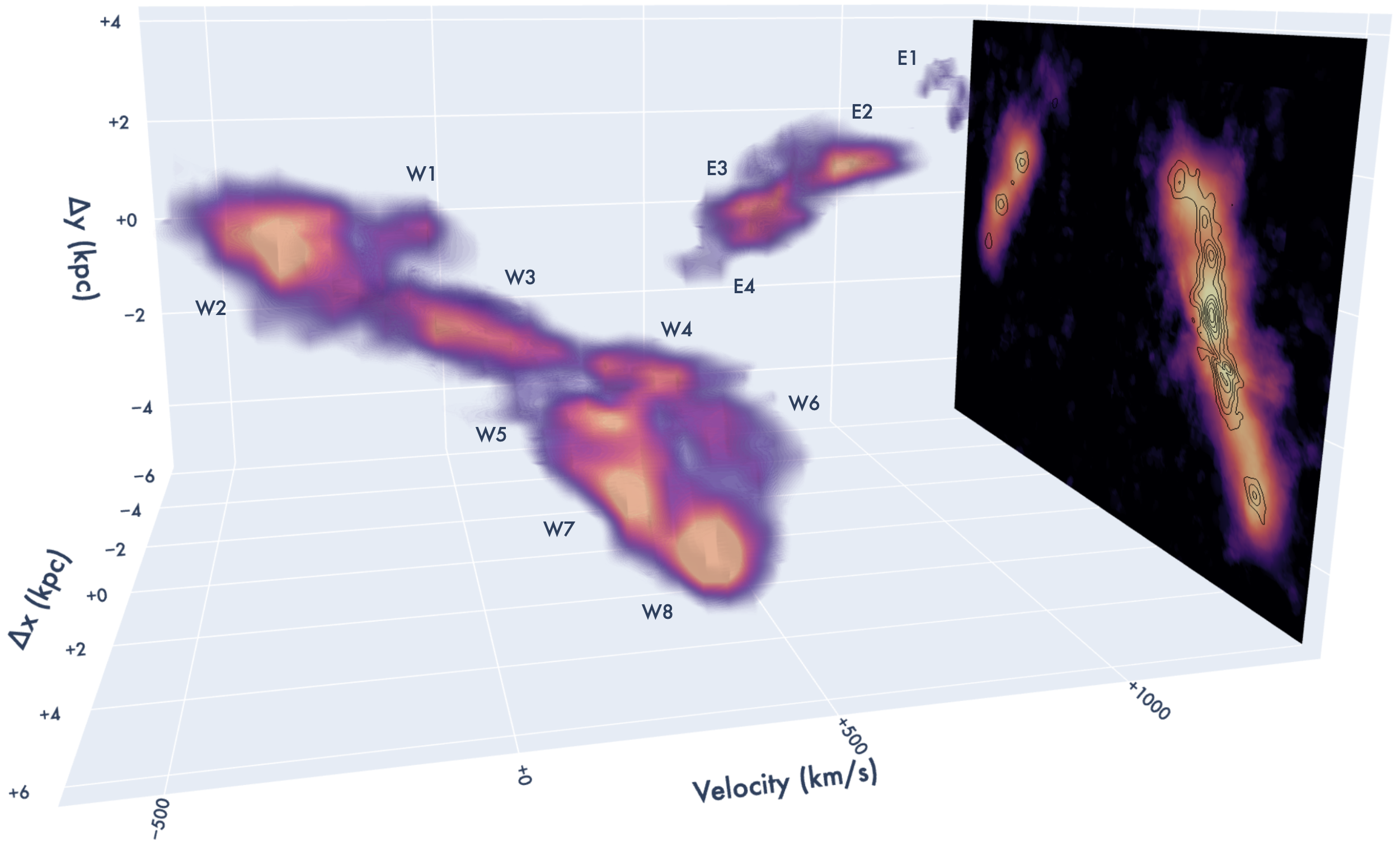}
\end{interactive}
\includegraphics[width=\textwidth]{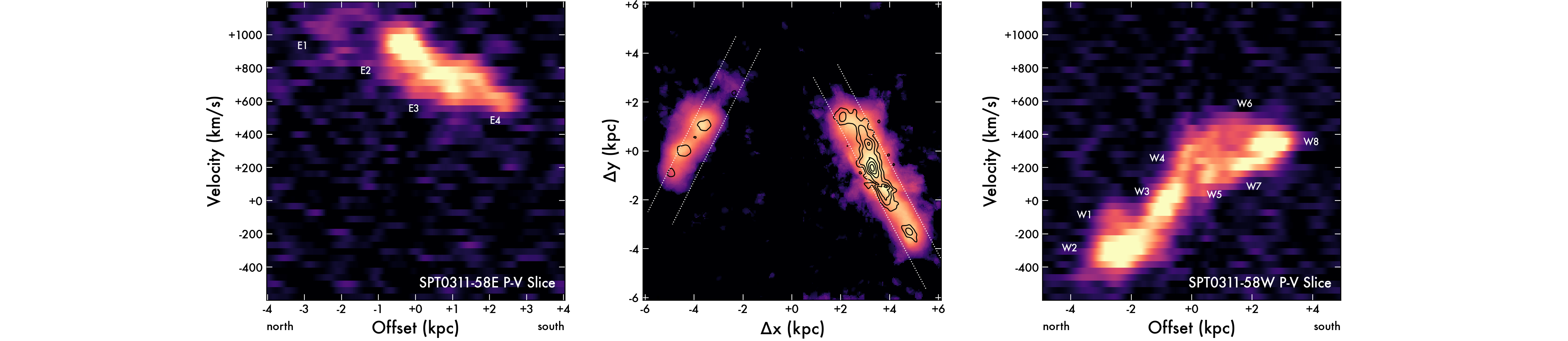}
\end{centering}
\caption{
Top: The \cii emission in SPT0311-58 is dominated by distinct clumps, distinguishable in this three-dimensional position-position-velocity rendering of the reconstructed data cube. The integrated \cii and dust continuum reconstructions are also shown, as in Fig.~\ref{fig:data}. An interactive version that allows the cube to be zoomed, panned, and rotated is available in the online journal and at \url{https://github.com/spt-smg/publicdata}.
Bottom: Position-velocity slices along SPT0311-58E~and~W, as indicated in the center panel, demonstrate that neither source is well-described by simple rotation curve models. Note that not all clumps are easily distinguishable in these 2-D projections of the 3-D cube.
}\label{fig:ciicube}
\end{figure*}

Our $\sim$100--350\,pc resolution data provide among the most detailed views of any reionization-era galaxy currently available, demonstrating that the SPT0311-58 system consists of at least a dozen smaller substructures that dominate the total dust and \cii emission. Importantly, no gravitational lensing `trickery' has artificially produced the clumps, as the structures identified by the algorithms are also readily apparent in the image-plane in this modestly-lensed system. We illustrate this correspondence by labeling the identified clumps in both the image and source planes in Fig.~\ref{fig:data}. Figure~\ref{fig:ciicube} demonstrates that the \cii emission is also highly clumpy in position-position-velocity space, with the smooth velocity-integrated emission a consequence of distinct clumps overlapping in the plane of the sky. The resolution and depth of our data (e.g. the continuum image reaches peak S/N$\approx$75) are sufficient to overcome limitations in past works that ruled out significant clumpy structure in lower-redshift DSFGs \citep[e.g.][]{hodge16,rujopakarn19,ivison20}.

Globally, the SPT0311-58 clumps are not arranged randomly in position-position-velocity space, but appear to trace two `strings' of clumps, one each in SPT0311-58W~and~E. While SPT0311-58E~and~W must be interacting, this raises the possibility that the clumps within each object have formed by fragmentation of a (very) massive pair of disks. The lower panels of Fig.~\ref{fig:ciicube} show position-velocity slices along the major axis of each object. Not all clumps are easily distinguished in these 2D projections of the 3D cube, although we have defined the slices to contain at least some of the \cii emission of all clumps. It is clear from Fig.~\ref{fig:ciicube} that neither SPT0311-58E~or~W is consistent with simple thin-disk rotation: not all clumps are co-linear in either source, and some clumps in each source do not follow the pattern expected from simple rotational models. This is in clear contrast to several recent works that claim rotational disk-like structure in lower-redshift DSFGs based on \cii kinematics \citep[e.g.][]{fraternali21,rizzo21} or dust continuum morphology alone \citep[e.g.][]{hodge19}. The P-V diagrams in Fig.~\ref{fig:ciicube} are clearly less well-defined, more disturbed, and more turbulent than the analogous diagrams in, e.g., \citealt{rizzo21}. Even if we ascribe the entirety of the structure seen in Fig.~\ref{fig:ciicube} to rotation, we would estimate an upper limit on the ratio of rotational-to-turbulent motions $V/\sigma \lesssim 2-3$ in both SPT0311-58 galaxies, well below that claimed by the aforementioned works. Whether due to the higher redshift or the much larger mass of the system compared to these objects, SPT0311-58 presents a clear counterexample to the recent narrative that early DSFGs are dominated by dynamically cold, secularly rotating systems.

We nevertheless expect that rotational motions are present at some level, for several reasons. Angular momentum conservation requires that merging structures orbit each other, producing `velocity gradients' superficially similar to rotation \citep[see, e.g.][for a similar case in a $z\sim5.7$ DSFG]{litke19}. However, it is implausible that a dozen merging subhalos should all be found within such close proximity yet still be distinctly identifiable given the short timescales expected for coalescence.  Moreover, local processes can distort the rotation curve from the classic patterns. For example, a position-velocity diagram of the M82 galactic disc shows evidence of a supernova-driven bubble qualitatively similar to the pattern seen in clumps W4--W8, though on spatial scales $\sim$10$\times$ smaller \citep{weiss99}. In all likelihood both disk fragmentation and gravitational interactions contribute to the clumpy structure of SPT0311-58. SPT0311-58E~and~W are separated by $<$10\,kpc on the sky and are clearly interacting, and at least some of the clumps in each source are likely distinct merging substructures. Other clumps may have fragmented from massive turbulent parent disks, but it is difficult to know the formation mechanism of any particular clump or ensemble of clumps.

Turning to the internal kinematics of the clumps, all clumps have velocity dispersions $\sigma = 70-120$\,\kms and circularized sizes $r_{\mathrm{circ,\cii}} = 0.7-1.3$\,kpc, measured from pseudo-moment-0 images that integrate the \cii emission of each clump along the spectral dimension.\footnote{These sizes are probably underestimated because the clump-finding algorithm does not allow pixels to belong to more than one clump i.e. clump overlap is ignored.} Only two clumps in SPT0311-58W, W3 and W4, show obvious internal velocity gradients that may suggest that the clumps themselves are rotating. Even for these clumps it is unclear whether rotation dominates, since these clumps are nearest the center of SPT0311-58W and are plausibly interacting. The remaining 10 clumps all appear to be dispersion-dominated at the current resolution. Future observations, especially those targeting a spectral line with more stringent excitation conditions to suppress emission from the diffuse gas, will be necessary to ascertain the internal kinematics of the SPT0311-58 clumps.

\begin{figure}
\begin{centering}
\includegraphics[width=\columnwidth]{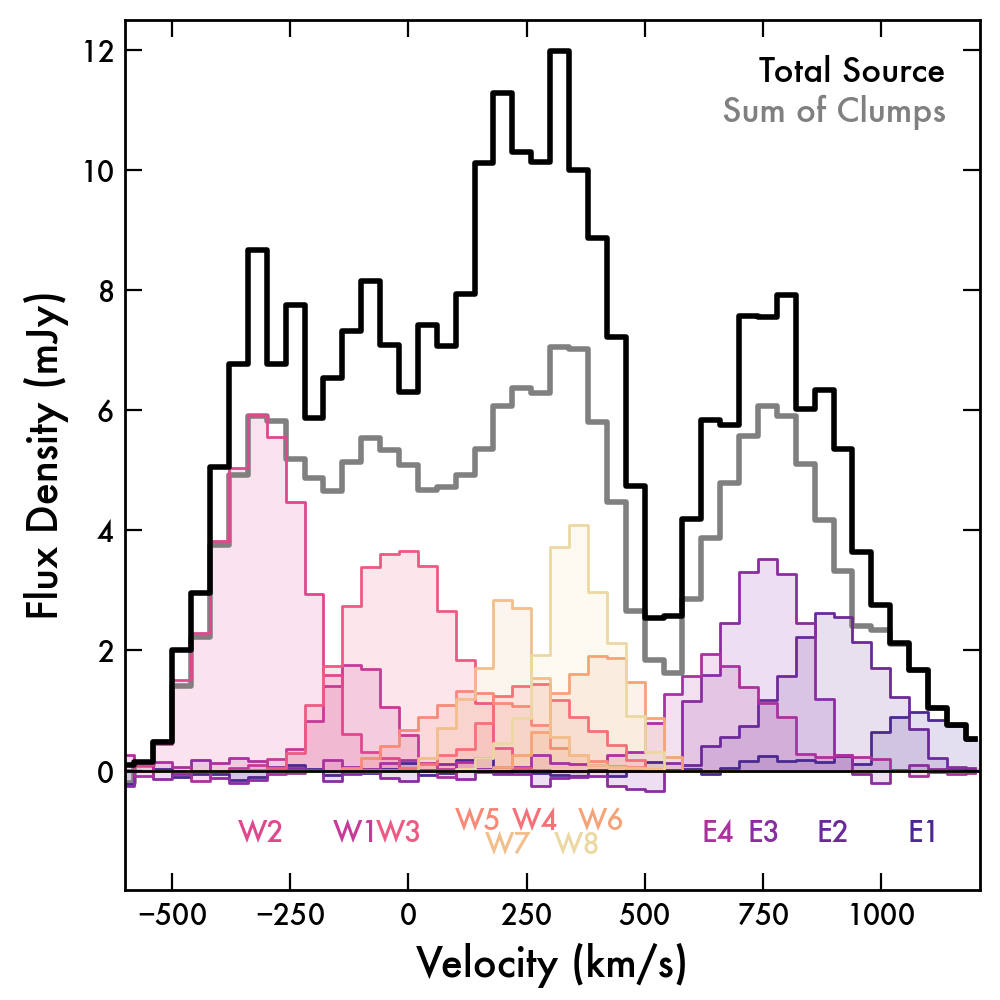}
\end{centering}
\caption{
\cii spectra of each identified clump (colors, labeled), the sum of the clump spectra (grey), and the total source-integrated emission (black). The clumps are kinematically (and spatially, Fig.~\ref{fig:ciicube}) distinct, and together comprise $\approx$70\% of the total \cii emission.
}\label{fig:ciispectra}
\end{figure}

\subsection{Enhanced Star Formation in z $\sim$ 7 Galaxy-Mass Clumps} \label{mdyn}

\begin{figure}
\begin{centering}
\includegraphics[width=\columnwidth]{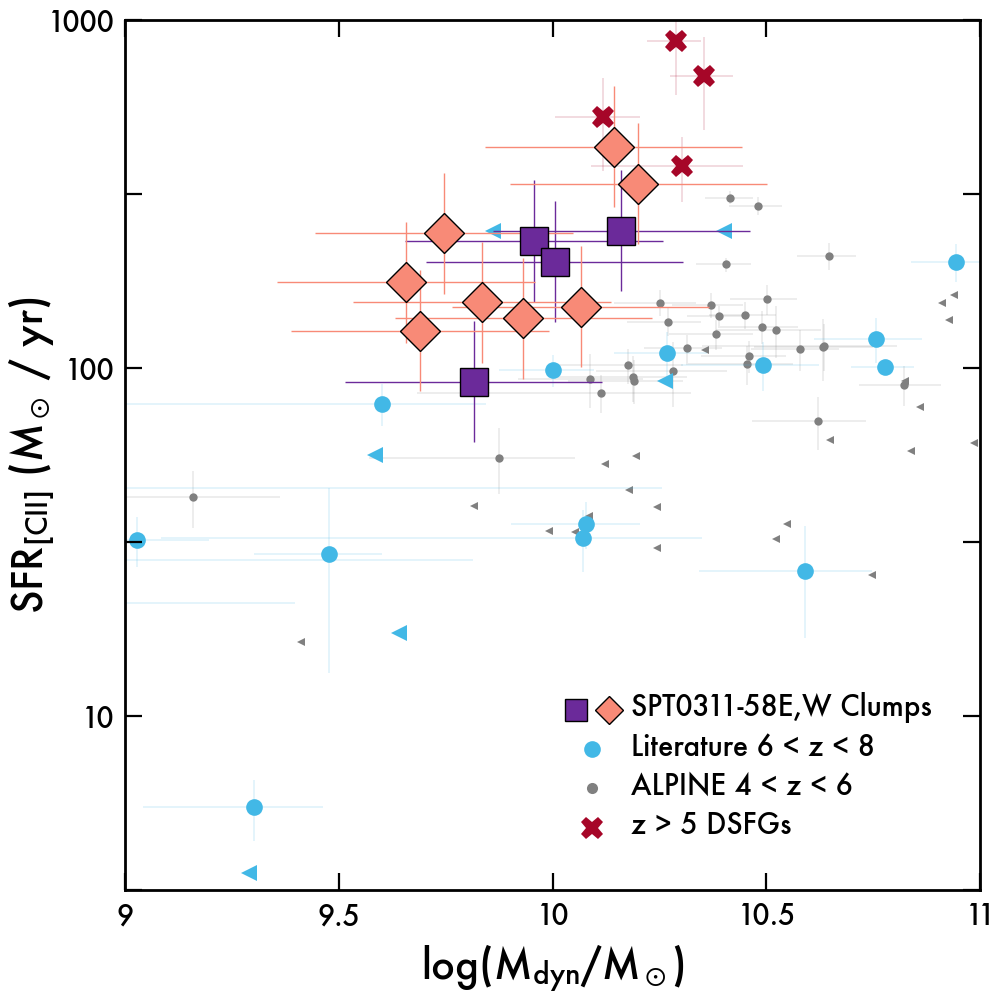}
\end{centering}
\caption{
The individual clumps identified in SPT0311-58 are similar in mass to `typical' co-eval massive galaxies, but show SFRs elevated by $\approx$3--5$\times$. However they formed, the clumps can be thought of as similar in scale to known reionization-era galaxies. All comparison objects also have spatially-resolved ALMA \cii observations; see Sec.~\ref{mdyn}.
}\label{fig:mdynsfr}
\end{figure}

We have shown that the clumps in SPT0311-58 are dispersion-dominated at the spatial resolution of the current data. Using the clump sizes and line widths from our clump-finding algorithm, we calculate simple dynamical mass estimates as $\Mdyn = \gamma \sigma^2 R / G$ with the clump velocity dispersion $\sigma$, radius $R$, and gravitational constant G. The dimensionless pre-factor $\gamma$ is intended to encapsulate the details of the clump kinematics and is typically assumed to lie in the range 2--7 (see \citealt{spilker15} and references therein). We adopt $\gamma=5$ for consistency with our subsequent analysis in Section~\ref{cprops}, appropriate for dispersion-dominated systems. We estimate uncertainties of at least a factor of 2 because of the difficulty in separating the emission from adjacent clumps with low contrast and the lack of knowledge of the detailed clump dynamics. We calculate clump SFRs using the \citet{delooze14} \cii-SFR relation; while calibrated at $z\sim0$, we expect this relation to provide reasonably good estimates at $z\gtrsim6$ as well \citep[e.g.][]{leung20}. We find generally consistent SFRs if we instead apportion the total \lir-based SFR among the clumps based on their continuum flux densities; \lir is globally well-constrained for these galaxies \citepalias{marrone18}. These SFRs are also likely uncertain by a factor of at least $\sim$2 for the same reasons as above, excluding systematic uncertainties in the \cii-SFR or \lir-SFR conversions, and any differences in the dust temperature between clumps within each source, which would redistribute the same well-measured \lir among the clumps differently.

Fig.~\ref{fig:mdynsfr} compares the SPT0311-58 clumps in \Mdyn--SFR with lower-redshift massive galaxies from the ALPINE survey \citep[$4<z<6$;][]{fujimoto20,bethermin20,faisst20}, a sample of (UV-selected) $6<z<8$ galaxies detected in \cii assembled from the literature \citep{willott15,knudsen16,pentericci16,bradac17,matthee17,carniani18,smit18,hashimoto19,matthee19,bakx20,harikane20,fujimoto21}, and $z>5$ IR-selected galaxies \citep{zavala18,litke19,spilker20a}. We use the same \cii-based SFR estimator for all samples, and the same \Mdyn calculation unless the original studies used a different method. The clump SFRs are clearly elevated compared to the UV-selected blank-field galaxies by $\approx$3--5$\times$, as are the other IR-selected $z>5$ objects. At some level this is a selection effect, since low-SFR galaxies by definition cannot enter the SPT survey  \citep{vieira13,spilker16,everett20}, but also reflects the long-known high star formation efficiency in DSFGs (SFE\,$\equiv$\,SFR$/\Mgas$, and to zeroth order $\Mdyn\sim\Mgas$; e.g. \citealt[][]{greve05,aravena16}). Interestingly, however, the masses of the individual SPT0311-58 clumps are very typical of $6<z<8$ `normal' galaxies selected from blank field imaging surveys. In other words, each clump individually can be thought of as similar in scale to the known population of high-redshift galaxies which together sum to create an extremely massive system, regardless of the physical origins of each individual clump.

\subsection{The Nature of the Massive Clumps} \label{cprops}

\begin{figure*}
\begin{centering}
\includegraphics[width=0.33\textwidth]{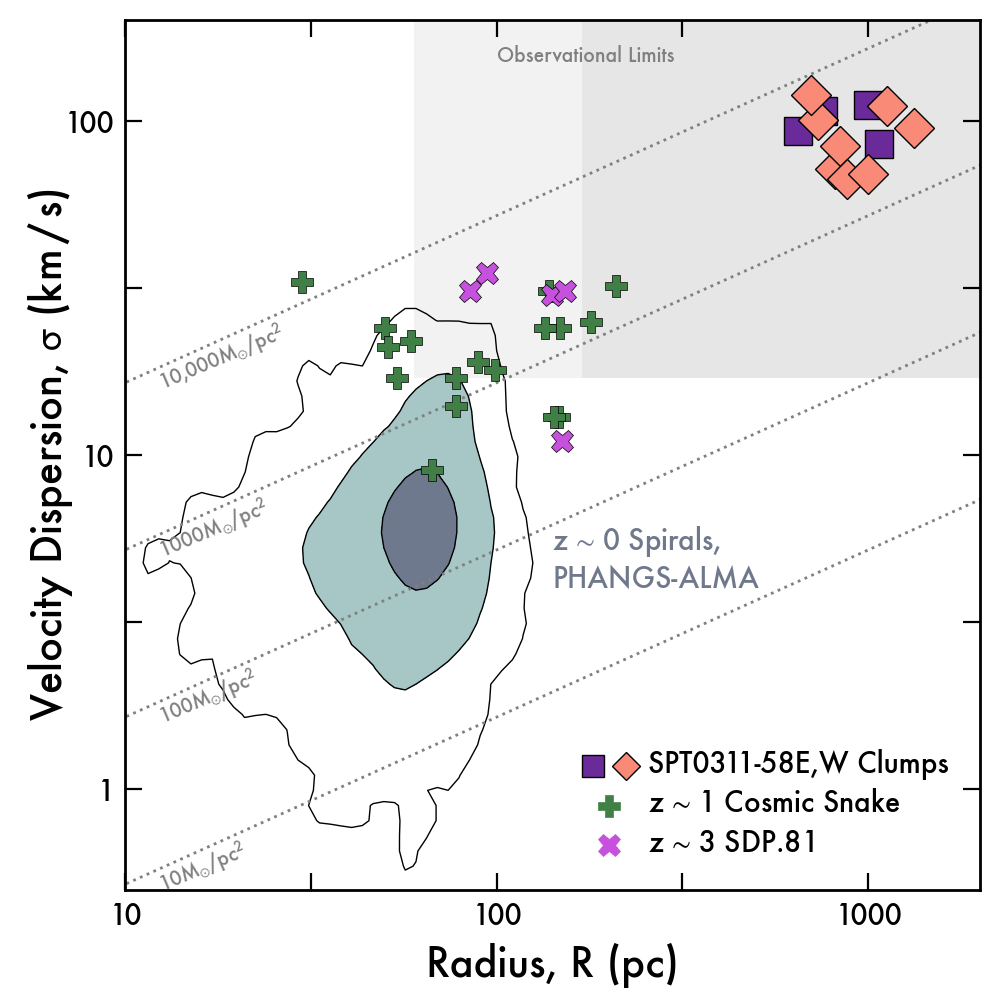}
\includegraphics[width=0.33\textwidth]{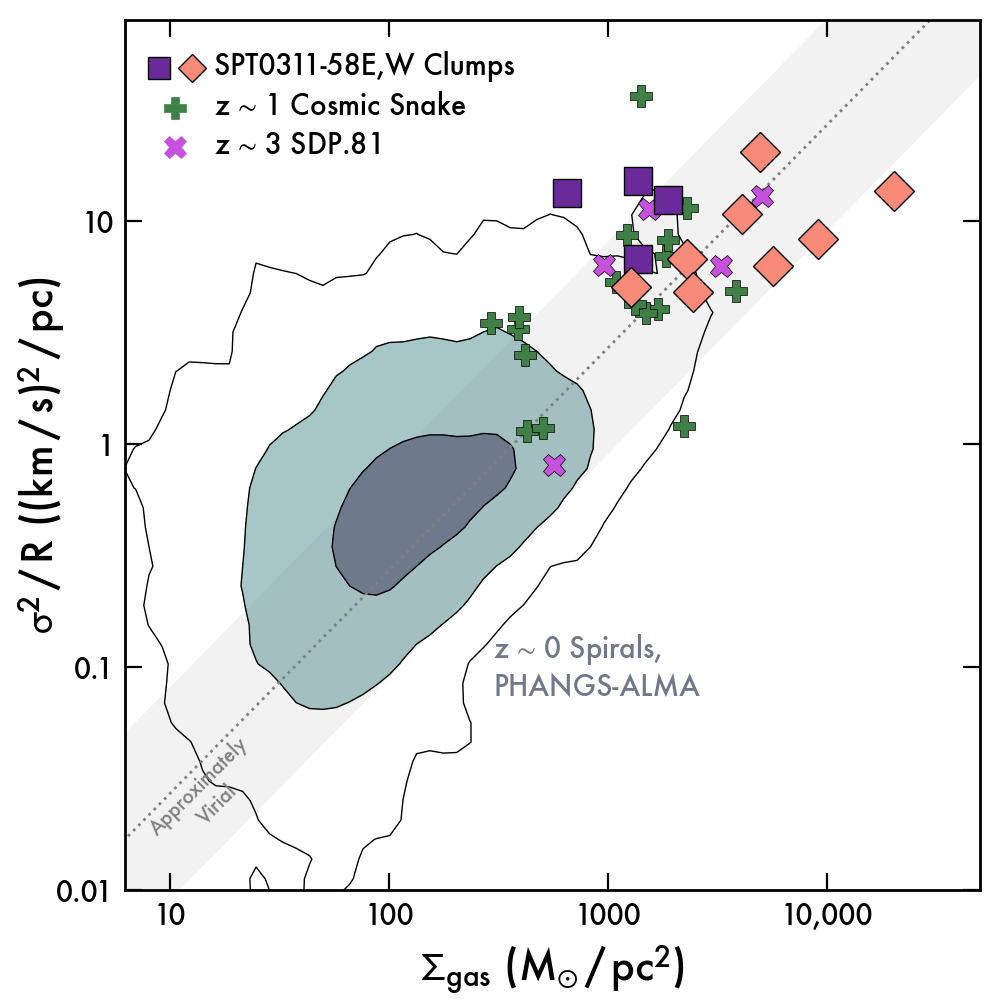}
\includegraphics[width=0.33\textwidth]{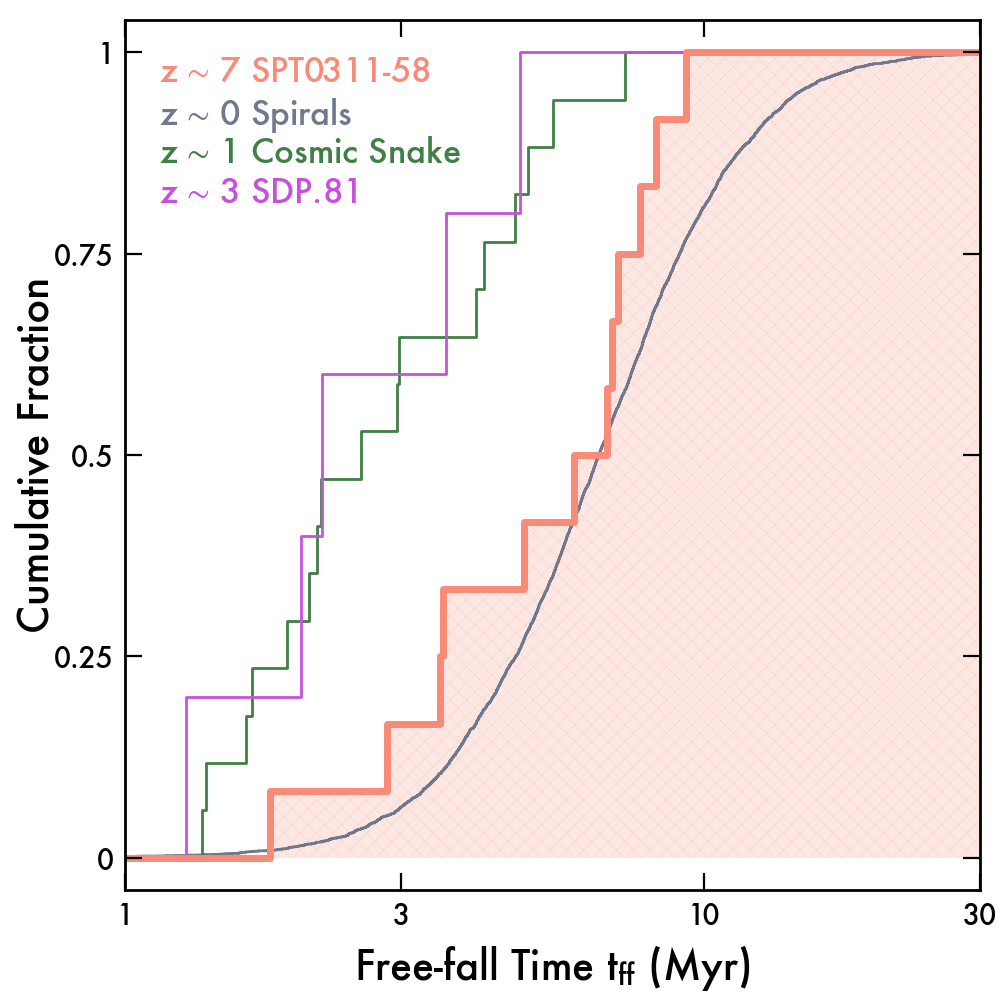}
\end{centering}
\caption{
Summary of SPT0311-58 clump properties in comparison to cloud-scale structures identified in nearby galaxies and two high-redshift lensed objects. Although more massive and larger in scale, the internal properties of the SPT0311-58 clumps are remarkably similar to the lower-redshift structures: they have similar gas surface densities as the z$\sim$1 and 3 clouds in the lensed objects (left), are approximately consistent with being virialized (center), and (assuming virialization) show a distribution of clump free-fall times similar to the lower-redshift clouds (right). In the left panel, the gray shaded regions show the approximate limits of our data in size and line width for both highly-magnified and unlensed regions of the source plane. Section~\ref{cprops} details the derivation of the dotted lines in the left and center panels.}\label{fig:cprops}
\end{figure*}

Finally, in this section we seek to gain more insight into the internal structure of the SPT0311-58 clumps from our $\sim$100--300\,pc resolution data. Our results are summarized by Figure~\ref{fig:cprops}. Fig.~\ref{fig:cprops} (left) shows the size--line width relation. We compare to a large sample of giant molecular clouds (GMCs) in nearby spiral galaxies from the PHANGS-ALMA survey \citep{rosolowsky21}, clumps identified in the lensed z$\sim$1 `Cosmic Snake' galaxy \citep{dessaugeszavadsky19}, and similar clumps in the lensed z$\sim$3 IR-selected galaxy SDP.81. The clumps in SPT0311-58 are clearly physically larger and more turbulent than the lower-redshift clump-scale structures. The gray shaded regions in Fig.~\ref{fig:cprops} (left) illustrate the approximate limits of our data in size and line width, with size limits corresponding to our best achievable resolution in high-magnification regions of the source plane as well as unlensed regions, and a line width limit corresponding to FWHM$=$40\,\kms, consistent with our recovery tests described in Section~\ref{clumpid}.

For virialized clouds with $M = 5 \sigma^2 R / G$, clumps at fixed mass surface density $\Sigma \propto M/R^2$ follow the dotted lines in Fig.~\ref{fig:cprops} (left). Although we examine this assumption next, we see from Fig.~\ref{fig:cprops} (left) that the implied mass surface densities of the SPT0311-58 clumps are comparable to those of other $z>1$ cloud complexes. However, given the observational limitations, it is clear that we only would have been able to identify perhaps a third of the clumps seen in the $z\sim1-3$ galaxies, with the remainder too small and/or too narrow to be present in our data. Nevertheless, we would have been sensitive to clumps $\sim$0.5\,dex smaller in size and/or $\sim$0.5\,dex narrower in line width, but no such clumps are identified -- in other words, none of the clumps we identify are close to the observational bounds of the data. Future observations with higher spatial resolution and/or sufficiently high S/N to allow reconstructions at finer spectral resolution will be required to determine if the clumps in SPT0311-58 break up in to yet smaller structures. The similarity in implied surface density suggests that, internally, the SPT0311-58 clumps may be akin to (greatly) `scaled-up' versions of similar structures in lower-redshift galaxies.

This is demonstrated further in Fig.~\ref{fig:cprops} (center), where we compare the clump gas surface densities to the velocity dispersion `normalized' by the clump size. Here we use the gas masses derived from radiative transfer modeling of the dust and CO emission in SPT0311-58 (\citealt{strandet17}, \citetalias{marrone18}, \citealt{jarugula21}), most similar to the CO-based masses available for the comparison clumps. We simply divide the total masses of SPT0311-58E~and~W among the clumps according to their continuum flux densities; this is equivalent to assuming a constant gas/dust mass ratio in each source, which seems reasonable. For clouds in virial equilibrium\footnote{Typically clouds within a factor of 3 of this relation are taken to be consistent with virialization \citep[e.g.][]{dessaugeszavadsky19,rosolowsky21}, which we also adopt here.} with mass dominated by the cold gas, we expect $\siggas \sim 370 \sigma^2 / R$ with \siggas in \Msol\,pc$^{-2}$, $\sigma$ in \kms, and R in pc. Nearly all SPT0311-58 clumps are consistent with the locus expected for virialized structures, suggesting that they are also consistent with being self-gravitating bound objects. We note that unresolved rotational motions would move the clumps downwards in this plot, making them more tightly bound. Similar to the large sample of $z\sim0$ clouds and the limited samples of high-redshift clumps, the substructures in SPT0311-58 appear to be approximately in virial equilibrium. Assuming virialization, we show the distribution of the gravitational free-fall time for the SPT0311-58 and literature comparison clumps in Fig.~\ref{fig:cprops} (right), following \citet{rosolowsky21} for the formulation of the free-fall time. The free-fall times of the SPT0311-58 clumps are remarkably similar to the distribution of the $z\sim0$ clumps. Equivalently, the volume density of the kpc-scale SPT0311-58 clumps is remarkably similar to the density of the $\sim$1000$\times$ smaller volume local clouds, since $t_{\mathrm{ff}} \propto 1/\sqrt{\rho}$. 

Taken together, Fig.~\ref{fig:cprops} suggests that the individual clumps in SPT0311-58 bear a striking resemblance to similar structures on smaller scales in lower-redshift galaxies. We have argued that gravitational interactions, both between SPT0311-58E~and~W and between individual clumps within each object, likely play a strong role in driving the overall structure of the galaxies (Section~\ref{kinematics}), even if some clumps formed from fragmentation of very massive parent disks. Nevertheless the internal properties of the clumps seem to be otherwise similar to those of GMC-scale structures that form in undisturbed disks. Future observations with even higher spatial resolution could measure more detailed internal kinematic structure within the clumps and/or resolve each clump into even smaller sub-structures. The very large amounts of cold gas contained in the clumps, combined with the high likelihood of interactions between clumps and between SPT0311-58E and SPT0311-58W, likely explains the very rapid star formation seen in these galaxies.

\section{Conclusions} \label{conclusions}

We have presented a detailed view of an extremely massive reionization-era system, using 0.07'' resolution observations of the dust and \cii emission in the $z=6.900$ SPT0311-58 system to probe the structure of the central galaxies on scales down to $\sim$100\,pc. The observations resolve the pair of central galaxies into at least a dozen clumps, each with turbulent velocity dispersion $\sigma = 70-120$\,\kms, size $r_{\mathrm{circ,\cii}} = 0.7-1.3$\,kpc, and mass $\log \Mdyn/\Msol \approx 10$. These galaxy-scale clumps are similar in mass to the population of currently-known reionization-era galaxies selected from rest-UV imaging surveys, but the SPT0311-58 clumps show SFRs elevated by about a factor of $\sim$3--5 compared to the UV-selected objects at similar mass. Globally, the chaotic and turbulent kinematics of SPT0311-58 present a striking contrast to recent claims of dynamically cold thin-disk rotation in some $z\sim4$ DSFGs. This could indicate a transition in the dominant DSFG kinematics between $z\sim7$ and $z\sim4$, although the short $\sim$800\,Myr interval between these epochs is only a few dynamical timescales. We speculate that some clumps may have formed from fragmentation of very massive parent disks while others are more likely merging substructures, but in any case the overall kinematics are highly chaotic and turbulent. Internally, the SPT0311-58 clumps bear a striking resemblance to greatly scaled-up versions of molecular cloud-scale structures identified in $z\sim0$ spirals and a pair of $z\sim1-3$ lensed galaxies; for example, the SPT0311-58 clumps show a similar distribution of free-fall times as a large sample of $z\sim0$ molecular clouds, implying similar volume densities as these clouds despite being $\gtrsim$10$\times$ larger in size.

Thus far our most detailed view of the SPT0311-58 system has come from the millimeter imaging presented here, tracing the cold dust and gas within the central galaxies. Upcoming observations with the \textit{James Webb Space Telescope} (GTO-1264, GO-1791) will also allow a similar-resolution view of the rest-UV/optical light and rest-optical nebular emission lines, allowing detailed constraints on the stellar contents and unobscured star formation in the central regions of this massive halo. Equally importantly, these observations will also probe larger distances into the halo surrounding the central galaxies, which will allow the selection of additional $z\sim7$ galaxies tracing the overdensity on larger scales. From the planned \textit{JWST} imaging and spectroscopy it may also be possible to place (loose) constraints on the very early $z\gtrsim10$ formation history of these galaxies, although the high dust obscuration will present challenges. Together with existing imaging from the \textit{Hubble Space Telescope} and our ALMA observations, we will soon have a comprehensive sub-kiloparsec scale view of the galaxies assembling within an extremely massive reionization-era dark matter halo.

\begin{acknowledgements}
JSS was supported in part by NASA Hubble Fellowship grant \#HF2-51446  awarded  by  the  Space  Telescope  Science  Institute,  which  is  operated  by  the  Association  of  Universities  for  Research  in  Astronomy,  Inc.,  for  NASA,  under  contract  NAS5-26555. 
MA acknowledges support from FONDECYT grant 1211951, CONICYT + PCI + INSTITUTO MAX PLANCK DE ASTRONOMIA MPG190030,  CONICYT+PCI+REDES 190194 and ANID BASAL project FB210003.
KCL, DPM, KP, and JDV acknowledge support from the US NSF under grants AST-1715213 and AST-1716127.
DN acknowledges support from the NSF via grant AST-1909153.

This paper makes use of the following ALMA data: ADS/JAO.ALMA\#2016.1.01293.S, ADS/JAO.ALMA\#2017.1.01423.S. ALMA is a partnership of ESO (representing its member states), NSF (USA) and NINS (Japan), together with NRC (Canada), MOST and ASIAA (Taiwan), and KASI (Republic of Korea), in cooperation with the Republic of Chile. The Joint ALMA Observatory is operated by ESO, AUI/NRAO and NAOJ. The National Radio Astronomy Observatory is a facility of the National Science Foundation operated under cooperative agreement by Associated Universities, Inc.
\end{acknowledgements}

\facility{ALMA}

\software{
CASA \citep{mcmullin07},
\texttt{ripples} \citep{hezaveh16},
\texttt{astropy} \citep{astropy18},
\texttt{matplotlib} \citep{hunter07}}

\bibliographystyle{aasjournal}

\begin{thebibliography}{}
\expandafter\ifx\csname natexlab\endcsname\relax\def\natexlab#1{#1}\fi
\providecommand{\url}[1]{\href{#1}{#1}}
\providecommand{\dodoi}[1]{doi:~\href{http://doi.org/#1}{\nolinkurl{#1}}}
\providecommand{\doeprint}[1]{\href{http://ascl.net/#1}{\nolinkurl{http://ascl.net/#1}}}
\providecommand{\doarXiv}[1]{\href{https://arxiv.org/abs/#1}{\nolinkurl{https://arxiv.org/abs/#1}}}

\bibitem[{Aravena {et~al.}(2016)Aravena, Spilker, Bethermin, Bothwell, Chapman,
  {de Breuck}, Furstenau, {G{\'o}nzalez-L{\'o}pez}, Greve, Litke, Ma, Malkan,
  Marrone, Murphy, Stark, Strandet, Vieira, Weiss, Welikala, Wong, \&
  Collier}]{aravena16}
Aravena, M., Spilker, J.~S., Bethermin, M., {et~al.} 2016, \mnras, 457, 4406,
  \dodoi{10.1093/mnras/stw275}

\bibitem[{{Astropy Collaboration} {et~al.}(2018){Astropy Collaboration},
  {Price-Whelan}, Sip{\H o}cz, G{\"u}nther, Lim, Crawford, Conseil, Shupe,
  Craig, Dencheva, Ginsburg, VanderPlas, Bradley, {P{\'e}rez-Su{\'a}rez}, {de
  Val-Borro}, Aldcroft, Cruz, Robitaille, Tollerud, Ardelean, Babej, Bach,
  Bachetti, Bakanov, Bamford, Barentsen, Barmby, Baumbach, Berry, Biscani,
  Boquien, Bostroem, Bouma, Brammer, Bray, Breytenbach, Buddelmeijer, Burke,
  Calderone, Cano~Rodr{\'i}guez, Cara, Cardoso, Cheedella, Copin, Corrales,
  Crichton, D'Avella, Deil, Depagne, Dietrich, Donath, Droettboom, Earl, Erben,
  Fabbro, Ferreira, Finethy, Fox, Garrison, Gibbons, Goldstein, Gommers, Greco,
  Greenfield, Groener, Grollier, Hagen, Hirst, Homeier, Horton, Hosseinzadeh,
  Hu, Hunkeler, Ivezi{\'c}, Jain, Jenness, Kanarek, Kendrew, Kern, Kerzendorf,
  Khvalko, King, Kirkby, Kulkarni, Kumar, Lee, Lenz, Littlefair, Ma, Macleod,
  Mastropietro, McCully, Montagnac, Morris, Mueller, Mumford, Muna, Murphy,
  Nelson, Nguyen, Ninan, N{\"o}the, Ogaz, Oh, Parejko, Parley, Pascual, Patil,
  Patil, Plunkett, Prochaska, Rastogi, Reddy~Janga, Sabater, Sakurikar,
  Seifert, Sherbert, {Sherwood-Taylor}, Shih, Sick, Silbiger, Singanamalla,
  Singer, Sladen, Sooley, Sornarajah, Streicher, Teuben, Thomas, Tremblay,
  Turner, Terr{\'o}n, {van Kerkwijk}, {de la Vega}, Watkins, Weaver, Whitmore,
  Woillez, Zabalza, \& {Astropy Contributors}}]{astropy18}
{Astropy Collaboration}, {Price-Whelan}, A.~M., Sip{\H o}cz, B.~M., {et~al.}
  2018, \aj, 156, 123, \dodoi{10.3847/1538-3881/aabc4f}

\bibitem[{Bakx {et~al.}(2020)Bakx, Tamura, Hashimoto, Inoue, Lee, Mawatari,
  Ota, Umehata, Zackrisson, Hatsukade, Kohno, Matsuda, Matsuo, Okamoto,
  Shibuya, Shimizu, Taniguchi, \& Yoshida}]{bakx20}
Bakx, T. J. L.~C., Tamura, Y., Hashimoto, T., {et~al.} 2020, \mnras, 493, 4294,
  \dodoi{10.1093/mnras/staa509}

\bibitem[{Berry(2015)}]{berry15}
Berry, D.~S. 2015, Astronomy and Computing, 10, 22,
  \dodoi{10.1016/j.ascom.2014.11.004}

\bibitem[{B{\'e}thermin {et~al.}(2020)B{\'e}thermin, Fudamoto, Ginolfi,
  Loiacono, Khusanova, Capak, Cassata, Faisst, Le~F{\`e}vre, Schaerer,
  Silverman, Yan, Amorin, Bardelli, Boquien, Cimatti, Davidzon,
  {Dessauges-Zavadsky}, Fujimoto, Gruppioni, Hathi, Ibar, Jones, Koekemoer,
  Lagache, Lemaux, Moreau, Oesch, Pozzi, Riechers, Talia, Toft, Vallini,
  Vergani, Zamorani, \& Zucca}]{bethermin20}
B{\'e}thermin, M., Fudamoto, Y., Ginolfi, M., {et~al.} 2020, \aap, 643, A2,
  \dodoi{10.1051/0004-6361/202037649}

\bibitem[{Brada{\v c} {et~al.}(2017)Brada{\v c}, {Garcia-Appadoo}, Huang,
  Vallini, Quinn~Finney, Hoag, Lemaux, Borello~Schmidt, Treu, Carilli,
  Dijkstra, Ferrara, Fontana, Jones, Ryan, Wagg, \& Gonzalez}]{bradac17}
Brada{\v c}, M., {Garcia-Appadoo}, D., Huang, K.-H., {et~al.} 2017, \apj, 836,
  L2, \dodoi{10.3847/2041-8213/836/1/L2}

\bibitem[{Carniani {et~al.}(2018)Carniani, Maiolino, Amorin, Pentericci,
  Pallottini, Ferrara, Willott, Smit, Matthee, Sobral, Santini, Castellano,
  De~Barros, Fontana, Grazian, \& Guaita}]{carniani18}
Carniani, S., Maiolino, R., Amorin, R., {et~al.} 2018, \mnras, 478, 1170,
  \dodoi{10.1093/mnras/sty1088}

\bibitem[{Cooray {et~al.}(2014)Cooray, Calanog, Wardlow, Bock, Bridge,
  Burgarella, Bussmann, Casey, Clements, Conley, Farrah, Fu, Gavazzi, Ivison,
  La~Porte, Lo~Faro, Ma, Magdis, Oliver, Osage, {P{\'e}rez-Fournon}, Riechers,
  Rigopoulou, Scott, Viero, \& Watson}]{cooray14}
Cooray, A., Calanog, J., Wardlow, J.~L., {et~al.} 2014, \apj, 790, 40,
  \dodoi{10.1088/0004-637X/790/1/40}

\bibitem[{De~Looze {et~al.}(2014)De~Looze, Cormier, Lebouteiller, Madden, Baes,
  Bendo, Boquien, Boselli, Clements, Cortese, Cooray, Galametz, Galliano,
  {Graci{\'a}-Carpio}, Isaak, Karczewski, Parkin, Pellegrini, {R{\'e}my-Ruyer},
  Spinoglio, Smith, \& Sturm}]{delooze14}
De~Looze, I., Cormier, D., Lebouteiller, V., {et~al.} 2014, \aap, 568, A62,
  \dodoi{10.1051/0004-6361/201322489}

\bibitem[{{Dessauges-Zavadsky} {et~al.}(2019){Dessauges-Zavadsky}, Richard,
  Combes, Schaerer, Rujopakarn, Mayer, Cava, Boone, Egami, Kneib,
  {P{\'e}rez-Gonz{\'a}lez}, Pfenniger, Rawle, Teyssier, \& {van der
  Werf}}]{dessaugeszavadsky19}
{Dessauges-Zavadsky}, M., Richard, J., Combes, F., {et~al.} 2019, Nature
  Astronomy, 3, 1115, \dodoi{10.1038/s41550-019-0874-0}

\bibitem[{Everett {et~al.}(2020)Everett, Zhang, Crawford, Vieira, Aravena,
  Archipley, Austermann, Benson, Bleem, Carlstrom, Chang, Chapman, Crites, {de
  Haan}, Dobbs, George, Halverson, Harrington, Holder, Holzapfel, Hrubes, Knox,
  Lee, {Luong-Van}, Mangian, Marrone, McMahon, Meyer, Mocanu, Mohr, Natoli,
  Padin, Pryke, Reichardt, Reuter, Ruhl, Sayre, Schaffer, Shirokoff, Spilker,
  Stalder, Staniszewski, Stark, Story, Switzer, Vanderlinde, Wei{\ss}, \&
  Williamson}]{everett20}
Everett, W.~B., Zhang, L., Crawford, T.~M., {et~al.} 2020, \apj, 900, 55,
  \dodoi{10.3847/1538-4357/ab9df7}

\bibitem[{Faisst {et~al.}(2020)Faisst, Schaerer, Lemaux, Oesch, Fudamoto,
  Cassata, B{\'e}thermin, Capak, Le~F{\`e}vre, Silverman, Yan, Ginolfi,
  Koekemoer, Morselli, Amor{\'i}n, Bardelli, Boquien, Brammer, Cimatti,
  {Dessauges-Zavadsky}, Fujimoto, Gruppioni, Hathi, Hemmati, Ibar, Jones,
  Khusanova, Loiacono, Pozzi, Talia, Tasca, Riechers, Rodighiero, Romano,
  Scoville, Toft, Vallini, Vergani, Zamorani, \& Zucca}]{faisst20}
Faisst, A.~L., Schaerer, D., Lemaux, B.~C., {et~al.} 2020, \apjs, 247, 61,
  \dodoi{10.3847/1538-4365/ab7ccd}

\bibitem[{Forrest {et~al.}(2020)Forrest, Marsan, Annunziatella, Wilson, Muzzin,
  Marchesini, Cooper, Chan, McConachie, Gomez, {Kado-Fong}, La~Barbera,
  {Lange-Vagle}, Nantais, Nonino, Saracco, Stefanon, \& {van der
  Burg}}]{forrest20b}
Forrest, B., Marsan, Z.~C., Annunziatella, M., {et~al.} 2020, \apj, 903, 47,
  \dodoi{10.3847/1538-4357/abb819}

\bibitem[{Fraternali {et~al.}(2021)Fraternali, Karim, Magnelli,
  {G{\'o}mez-Guijarro}, {Jim{\'e}nez-Andrade}, \& Posses}]{fraternali21}
Fraternali, F., Karim, A., Magnelli, B., {et~al.} 2021, \aap, 647, A194,
  \dodoi{10.1051/0004-6361/202039807}

\bibitem[{Fujimoto {et~al.}(2020)Fujimoto, Silverman, Bethermin, Ginolfi,
  Jones, Le~F{\`e}vre, {Dessauges-Zavadsky}, Rujopakarn, Faisst, Fudamoto,
  Cassata, Morselli, Maiolino, Schaerer, Capak, Yan, Vallini, Toft, Loiacono,
  Zamorani, Talia, Narayanan, Hathi, Lemaux, Boquien, Amorin, Ibar, Koekemoer,
  {M{\'e}ndez-Hern{\'a}ndez}, Bardelli, Vergani, Zucca, Romano, \&
  Cimatti}]{fujimoto20}
Fujimoto, S., Silverman, J.~D., Bethermin, M., {et~al.} 2020, \apj, 900, 1,
  \dodoi{10.3847/1538-4357/ab94b3}

\bibitem[{Fujimoto {et~al.}(2021)Fujimoto, Oguri, Brammer, Yoshimura, Laporte,
  {Gonz{\'a}lez-L{\'o}pez}, Caminha, Kohno, Zitrin, Richard, Ouchi, Bauer,
  Smail, Hatsukade, Ono, Kokorev, Umehata, Schaerer, Knudsen, Sun, Magdis,
  Valentino, Ao, Toft, {Dessauges-Zavadsky}, Shimasaku, Caputi, Kusakabe,
  {Morokuma-Matsui}, Shotaro, Egami, Lee, Rawle, \& Espada}]{fujimoto21}
Fujimoto, S., Oguri, M., Brammer, G., {et~al.} 2021, \apj, 911, 99,
  \dodoi{10.3847/1538-4357/abd7ec}

\bibitem[{Glazebrook {et~al.}(2017)Glazebrook, Schreiber, Labb{\'e},
  Nanayakkara, Kacprzak, Oesch, Papovich, Spitler, Straatman, Tran, \&
  Yuan}]{glazebrook17}
Glazebrook, K., Schreiber, C., Labb{\'e}, I., {et~al.} 2017, \nat, 544, 71,
  \dodoi{10.1038/nature21680}

\bibitem[{Greve {et~al.}(2005)Greve, Bertoldi, Smail, Neri, Chapman, Blain,
  Ivison, Genzel, Omont, Cox, Tacconi, \& Kneib}]{greve05}
Greve, T.~R., Bertoldi, F., Smail, I., {et~al.} 2005, \mnras, 359, 1165,
  \dodoi{10.1111/j.1365-2966.2005.08979.x}

\bibitem[{Harikane {et~al.}(2020)Harikane, Ouchi, Inoue, Matsuoka, Tamura,
  Bakx, Fujimoto, Moriwaki, Ono, Nagao, Tadaki, Kojima, Shibuya, Egami,
  Ferrara, Gallerani, Hashimoto, Kohno, Matsuda, Matsuo, Pallottini, Sugahara,
  \& Vallini}]{harikane20}
Harikane, Y., Ouchi, M., Inoue, A.~K., {et~al.} 2020, \apj, 896, 93,
  \dodoi{10.3847/1538-4357/ab94bd}

\bibitem[{Hashimoto {et~al.}(2019)Hashimoto, Inoue, Mawatari, Tamura, Matsuo,
  Furusawa, Harikane, Shibuya, Knudsen, Kohno, Ono, Zackrisson, Okamoto,
  Kashikawa, Oesch, Ouchi, Ota, Shimizu, Taniguchi, Umehata, \&
  Watson}]{hashimoto19}
Hashimoto, T., Inoue, A.~K., Mawatari, K., {et~al.} 2019, \pasj, 71, 71,
  \dodoi{10.1093/pasj/psz049}

\bibitem[{Hezaveh {et~al.}(2016)Hezaveh, Dalal, Marrone, Mao, Morningstar, Wen,
  Blandford, Carlstrom, Fassnacht, Holder, Kemball, Marshall, Murray,
  Perreault~Levasseur, Vieira, \& Wechsler}]{hezaveh16}
Hezaveh, Y.~D., Dalal, N., Marrone, D.~P., {et~al.} 2016, \apj, 823, 37,
  \dodoi{10.3847/0004-637X/823/1/37}

\bibitem[{Hodge {et~al.}(2016)Hodge, Swinbank, Simpson, Smail, Walter,
  Alexander, Bertoldi, Biggs, Brandt, Chapman, Chen, Coppin, Cox, Dannerbauer,
  Edge, Greve, Ivison, Karim, Knudsen, Menten, Rix, Schinnerer, Wardlow, Weiss,
  \& {van der Werf}}]{hodge16}
Hodge, J.~A., Swinbank, A.~M., Simpson, J.~M., {et~al.} 2016, \apj, 833, 103,
  \dodoi{10.3847/1538-4357/833/1/103}

\bibitem[{Hodge {et~al.}(2019)Hodge, Smail, Walter, {da Cunha}, Swinbank,
  Rybak, Venemans, Brandt, Calistro~Rivera, Chapman, Chen, Cox, Dannerbauer,
  Decarli, Greve, Knudsen, Menten, Schinnerer, Simpson, {van der Werf},
  Wardlow, \& Weiss}]{hodge19}
Hodge, J.~A., Smail, I., Walter, F., {et~al.} 2019, \apj, 876, 130,
  \dodoi{10.3847/1538-4357/ab1846}

\bibitem[{Hunter(2007)}]{hunter07}
Hunter, J.~D. 2007, Computing in Science and Engineering, 9, 90,
  \dodoi{10.1109/MCSE.2007.55}

\bibitem[{Ivison {et~al.}(2020)Ivison, Richard, Biggs, Zwaan, Falgarone,
  Arumugam, {van der Werf}, \& Rujopakarn}]{ivison20}
Ivison, R.~J., Richard, J., Biggs, A.~D., {et~al.} 2020, \mnras,
  \dodoi{10.1093/mnrasl/slaa046}

\bibitem[{Ivison {et~al.}(1998)Ivison, Smail, Le~Borgne, Blain, Kneib,
  Bezecourt, Kerr, \& Davies}]{ivison98}
Ivison, R.~J., Smail, I., Le~Borgne, J.~F., {et~al.} 1998, \mnras, 298, 583,
  \dodoi{10.1046/j.1365-8711.1998.01677.x}

\bibitem[{Jarugula {et~al.}(2021)Jarugula, Vieira, Weiss, Spilker, Aravena,
  Archipley, B{\'e}thermin, Chapman, Dong, Greve, Harrington, Hayward, Hezaveh,
  Hill, Litke, Malkan, Marrone, Narayanan, Phadke, Reuter, \&
  Rotermund}]{jarugula21}
Jarugula, S., Vieira, J.~D., Weiss, A., {et~al.} 2021, \apj, 921, 97,
  \dodoi{10.3847/1538-4357/ac21db}

\bibitem[{Knudsen {et~al.}(2016)Knudsen, Richard, Kneib, Jauzac, Cl{\'e}ment,
  Drouart, Egami, \& Lindroos}]{knudsen16}
Knudsen, K.~K., Richard, J., Kneib, J.-P., {et~al.} 2016, \mnras, 462, L6,
  \dodoi{10.1093/mnrasl/slw114}

\bibitem[{Leung {et~al.}(2020)Leung, Olsen, Somerville, Dav{\'e}, Greve,
  Hayward, Narayanan, \& Popping}]{leung20}
Leung, T. K.~D., Olsen, K.~P., Somerville, R.~S., {et~al.} 2020, \apj, 905,
  102, \dodoi{10.3847/1538-4357/abc25e}

\bibitem[{Litke {et~al.}(2019)Litke, Marrone, Spilker, Aravena, B{\'e}thermin,
  Chapman, Chen, {de Breuck}, Dong, Gonzalez, Greve, Hayward, Hezaveh,
  Jarugula, Ma, Morningstar, Narayanan, Phadke, Reuter, Vieira, \&
  Weiss}]{litke19}
Litke, K.~C., Marrone, D.~P., Spilker, J.~S., {et~al.} 2019, \apj, 870, 80,
  \dodoi{10.3847/1538-4357/aaf057}

\bibitem[{Marrone {et~al.}(2018)Marrone, Spilker, Hayward, Vieira, Aravena,
  Ashby, Bayliss, B{\'e}thermin, Brodwin, Bothwell, Carlstrom, Chapman, Chen,
  Crawford, Cunningham, De~Breuck, Fassnacht, Gonzalez, Greve, Hezaveh,
  Lacaille, Litke, Lower, Ma, Malkan, Miller, Morningstar, Murphy, Narayanan,
  Phadke, Rotermund, Sreevani, Stalder, Stark, Strandet, Tang, \&
  Wei{\ss}}]{marrone18}
Marrone, D.~P., Spilker, J.~S., Hayward, C.~C., {et~al.} 2018, \nat, 553, 51,
  \dodoi{10.1038/nature24629}

\bibitem[{Matthee {et~al.}(2017)Matthee, Sobral, Boone, R{\"o}ttgering,
  Schaerer, Girard, Pallottini, Vallini, Ferrara, Darvish, \&
  Mobasher}]{matthee17}
Matthee, J., Sobral, D., Boone, F., {et~al.} 2017, \apj, 851, 145,
  \dodoi{10.3847/1538-4357/aa9931}

\bibitem[{Matthee {et~al.}(2019)Matthee, Sobral, Boogaard, R{\"o}ttgering,
  Vallini, Ferrara, {Paulino-Afonso}, Boone, Schaerer, \& Mobasher}]{matthee19}
Matthee, J., Sobral, D., Boogaard, L.~A., {et~al.} 2019, \apj, 881, 124,
  \dodoi{10.3847/1538-4357/ab2f81}

\bibitem[{McMullin {et~al.}(2007)McMullin, Waters, Schiebel, Young, \&
  Golap}]{mcmullin07}
McMullin, J.~P., Waters, B., Schiebel, D., Young, W., \& Golap, K. 2007, 376,
  127.
\newblock \url{http://adsabs.harvard.edu/abs/2007ASPC..376..127M}

\bibitem[{Newman {et~al.}(2018)Newman, Belli, Ellis, \& Patel}]{newman18}
Newman, A.~B., Belli, S., Ellis, R.~S., \& Patel, S.~G. 2018, \apj, 862, 126,
  \dodoi{10.3847/1538-4357/aacd4f}

\bibitem[{Pentericci {et~al.}(2016)Pentericci, Carniani, Castellano, Fontana,
  Maiolino, Guaita, Vanzella, Grazian, Santini, Yan, Cristiani, Conselice,
  Giavalisco, Hathi, \& Koekemoer}]{pentericci16}
Pentericci, L., Carniani, S., Castellano, M., {et~al.} 2016, \apj, 829, L11,
  \dodoi{10.3847/2041-8205/829/1/L11}

\bibitem[{{Planck Collaboration} {et~al.}(2016){Planck Collaboration}, Ade,
  Aghanim, Arnaud, Ashdown, Aumont, Baccigalupi, Banday, Barreiro, Bartlett,
  Bartolo, Battaner, Battye, Benabed, Beno{\^i}t, {Benoit-L{\'e}vy}, Bernard,
  Bersanelli, Bielewicz, Bock, Bonaldi, Bonavera, Bond, Borrill, Bouchet,
  Boulanger, Bucher, Burigana, Butler, Calabrese, Cardoso, Catalano, Challinor,
  Chamballu, Chary, Chiang, Chluba, Christensen, Church, Clements, Colombi,
  Colombo, Combet, Coulais, Crill, Curto, Cuttaia, Danese, Davies, Davis, {de
  Bernardis}, {de Rosa}, {de Zotti}, Delabrouille, D{\'e}sert, Di~Valentino,
  Dickinson, Diego, Dolag, Dole, Donzelli, Dor{\'e}, Douspis, Ducout, Dunkley,
  Dupac, Efstathiou, Elsner, En{\ss}lin, Eriksen, Farhang, Fergusson, Finelli,
  Forni, Frailis, Fraisse, Franceschi, Frejsel, Galeotta, Galli, Ganga,
  Gauthier, Gerbino, Ghosh, Giard, {Giraud-H{\'e}raud}, Giusarma, Gjerl{\o}w,
  {Gonz{\'a}lez-Nuevo}, G{\'o}rski, Gratton, Gregorio, Gruppuso, Gudmundsson,
  Hamann, Hansen, Hanson, Harrison, Helou, {Henrot-Versill{\'e}},
  {Hern{\'a}ndez-Monteagudo}, Herranz, Hildebrandt, Hivon, Hobson, Holmes,
  Hornstrup, Hovest, Huang, Huffenberger, Hurier, Jaffe, Jaffe, Jones, Juvela,
  Keih{\"a}nen, Keskitalo, Kisner, Kneissl, Knoche, Knox, Kunz, {Kurki-Suonio},
  Lagache, L{\"a}hteenm{\"a}ki, Lamarre, Lasenby, Lattanzi, Lawrence, Leahy,
  Leonardi, Lesgourgues, Levrier, Lewis, Liguori, Lilje, {Linden-V{\o}rnle},
  {L{\'o}pez-Caniego}, Lubin, {Mac{\'i}as-P{\'e}rez}, Maggio, Maino, Mandolesi,
  Mangilli, Marchini, Maris, Martin, Martinelli, {Mart{\'i}nez-Gonz{\'a}lez},
  Masi, Matarrese, McGehee, Meinhold, Melchiorri, Melin, Mendes, Mennella,
  Migliaccio, Millea, Mitra, {Miville-Desch{\^e}nes}, Moneti, Montier,
  Morgante, Mortlock, Moss, Munshi, Murphy, Naselsky, Nati, Natoli,
  Netterfield, {N{\o}rgaard-Nielsen}, Noviello, Novikov, Novikov, Oxborrow,
  Paci, Pagano, Pajot, Paladini, Paoletti, Partridge, Pasian, Patanchon,
  Pearson, Perdereau, Perotto, Perrotta, Pettorino, Piacentini, Piat,
  Pierpaoli, Pietrobon, Plaszczynski, Pointecouteau, Polenta, Popa, Pratt,
  Pr{\'e}zeau, Prunet, Puget, Rachen, Reach, Rebolo, Reinecke, Remazeilles,
  Renault, Renzi, Ristorcelli, Rocha, Rosset, Rossetti, Roudier, {Rouill{\'e}
  d'Orfeuil}, {Rowan-Robinson}, {Rubi{\~n}o-Mart{\'i}n}, Rusholme, Said,
  Salvatelli, Salvati, Sandri, Santos, Savelainen, Savini, Scott, Seiffert,
  Serra, Shellard, Spencer, Spinelli, Stolyarov, Stompor, Sudiwala, Sunyaev,
  Sutton, {Suur-Uski}, Sygnet, Tauber, Terenzi, Toffolatti, Tomasi, Tristram,
  Trombetti, Tucci, Tuovinen, T{\"u}rler, Umana, Valenziano, Valiviita,
  Van~Tent, Vielva, Villa, Wade, Wandelt, Wehus, White, White, Wilkinson, Yvon,
  Zacchei, \& Zonca}]{planck15}
{Planck Collaboration}, Ade, P. A.~R., Aghanim, N., {et~al.} 2016, \aap, 594,
  A13, \dodoi{10.1051/0004-6361/201525830}

\bibitem[{Rizzo {et~al.}(2021)Rizzo, Vegetti, Fraternali, Stacey, \&
  Powell}]{rizzo21}
Rizzo, F., Vegetti, S., Fraternali, F., Stacey, H.~R., \& Powell, D. 2021,
  \mnras, 507, 3952, \dodoi{10.1093/mnras/stab2295}

\bibitem[{Rosolowsky {et~al.}(2021)Rosolowsky, Hughes, Leroy, Sun, Querejeta,
  Schruba, Usero, Herrera, Liu, Pety, Saito, Be{\v s}li{\'c}, Bigiel, Blanc,
  Chevance, Dale, Deger, Faesi, Glover, Henshaw, Klessen, Kruijssen, Larson,
  Lee, Meidt, Mok, Schinnerer, Thilker, \& Williams}]{rosolowsky21}
Rosolowsky, E., Hughes, A., Leroy, A.~K., {et~al.} 2021, \mnras, 502, 1218,
  \dodoi{10.1093/mnras/stab085}

\bibitem[{Rujopakarn {et~al.}(2019)Rujopakarn, Daddi, Rieke, Puglisi, Schramm,
  {P{\'e}rez-Gonz{\'a}lez}, Magdis, Alberts, Bournaud, Elbaz, Franco,
  Kawinwanichakij, Kohno, Narayanan, Silverman, Wang, \&
  Williams}]{rujopakarn19}
Rujopakarn, W., Daddi, E., Rieke, G.~H., {et~al.} 2019, \apj, 882, 107,
  \dodoi{10.3847/1538-4357/ab3791}

\bibitem[{Smit {et~al.}(2018)Smit, Bouwens, Carniani, Oesch, Labb{\'e},
  Illingworth, {van der Werf}, Bradley, Gonzalez, Hodge, Holwerda, Maiolino, \&
  Zheng}]{smit18}
Smit, R., Bouwens, R.~J., Carniani, S., {et~al.} 2018, \nat, 553, 178,
  \dodoi{10.1038/nature24631}

\bibitem[{Spilker {et~al.}(2015)Spilker, Aravena, Marrone, B{\'e}thermin,
  Bothwell, Carlstrom, Chapman, Collier, {de Breuck}, Fassnacht, Galvin,
  Gonzalez, {Gonz{\'a}lez-L{\'o}pez}, Grieve, Hezaveh, Ma, Malkan, O'Brien,
  Rotermund, Strandet, Vieira, Weiss, \& Wong}]{spilker15}
Spilker, J.~S., Aravena, M., Marrone, D.~P., {et~al.} 2015, \apj, 811, 124,
  \dodoi{10.1088/0004-637X/811/2/124}

\bibitem[{Spilker {et~al.}(2016)Spilker, Marrone, Aravena, B{\'e}thermin,
  Bothwell, Carlstrom, Chapman, Crawford, {de Breuck}, Fassnacht, Gonzalez,
  Greve, Hezaveh, Litke, Ma, Malkan, Rotermund, Strandet, Vieira, Weiss, \&
  Welikala}]{spilker16}
Spilker, J.~S., Marrone, D.~P., Aravena, M., {et~al.} 2016, \apj, 826, 112,
  \dodoi{10.3847/0004-637X/826/2/112}

\bibitem[{Spilker {et~al.}(2020)Spilker, Phadke, Aravena, B{\'e}thermin,
  Chapman, Dong, Gonzalez, Hayward, Hezaveh, Jarugula, Litke, Malkan, Marrone,
  Narayanan, Reuter, Vieira, \& Weiss}]{spilker20a}
Spilker, J.~S., Phadke, K.~A., Aravena, M., {et~al.} 2020, \apj, 905, 85,
  \dodoi{10.3847/1538-4357/abc47f}

\bibitem[{Straatman {et~al.}(2016)Straatman, Spitler, Quadri, Labb{\'e},
  Glazebrook, Persson, Papovich, Tran, Brammer, Cowley, Tomczak, Nanayakkara,
  Alcorn, Allen, Broussard, {van Dokkum}, Forrest, {van Houdt}, Kacprzak,
  Kawinwanichakij, Kelson, Lee, McCarthy, Mehrtens, Monson, Murphy, Rees,
  Tilvi, \& Whitaker}]{straatman16}
Straatman, C. M.~S., Spitler, L.~R., Quadri, R.~F., {et~al.} 2016, \apj, 830,
  51, \dodoi{10.3847/0004-637X/830/1/51}

\bibitem[{Strandet {et~al.}(2017)Strandet, Weiss, De~Breuck, Marrone, Vieira,
  Aravena, Ashby, B{\'e}thermin, Bothwell, Bradford, Carlstrom, Chapman,
  Cunningham, Chen, Fassnacht, Gonzalez, Greve, Gullberg, Hayward, Hezaveh,
  Litke, Ma, Malkan, Menten, Miller, Murphy, Narayanan, Phadke, Rotermund,
  Spilker, \& Sreevani}]{strandet17}
Strandet, M.~L., Weiss, A., De~Breuck, C., {et~al.} 2017, \apjl, 842, L15,
  \dodoi{10.3847/2041-8213/aa74b0}

\bibitem[{Tacconi {et~al.}(2008)Tacconi, Genzel, Smail, Neri, Chapman, Ivison,
  Blain, Cox, Omont, Bertoldi, Greve, F{\"o}rster~Schreiber, Genel, Lutz,
  Swinbank, Shapley, Erb, Cimatti, Daddi, \& Baker}]{tacconi08}
Tacconi, L.~J., Genzel, R., Smail, I., {et~al.} 2008, \apj, 680, 246,
  \dodoi{10.1086/587168}

\bibitem[{Toft {et~al.}(2017)Toft, Zabl, Richard, Gallazzi, Zibetti, Prescott,
  Grillo, Man, Lee, {G{\'o}mez-Guijarro}, Stockmann, Magdis, \&
  Steinhardt}]{toft17}
Toft, S., Zabl, J., Richard, J., {et~al.} 2017, \nat, 546, 510,
  \dodoi{10.1038/nature22388}

\bibitem[{Vieira {et~al.}(2013)Vieira, Marrone, Chapman, De~Breuck, Hezaveh,
  Wei{$\beta$}, Aguirre, Aird, Aravena, Ashby, Bayliss, Benson, Biggs, Bleem,
  Bock, Bothwell, Bradford, Brodwin, Carlstrom, Chang, Crawford, Crites, {de
  Haan}, Dobbs, Fomalont, Fassnacht, George, Gladders, Gonzalez, Greve,
  Gullberg, Halverson, High, Holder, Holzapfel, Hoover, Hrubes, Hunter,
  Keisler, Lee, Leitch, Lueker, {Luong-van}, Malkan, McIntyre, McMahon, Mehl,
  Menten, Meyer, Mocanu, Murphy, Natoli, Padin, Plagge, Reichardt, Rest, Ruel,
  Ruhl, Sharon, Schaffer, Shaw, Shirokoff, Spilker, Stalder, Staniszewski,
  Stark, Story, Vanderlinde, Welikala, \& Williamson}]{vieira13}
Vieira, J.~D., Marrone, D.~P., Chapman, S.~C., {et~al.} 2013, \nat, 495, 344,
  \dodoi{10.1038/nature12001}

\bibitem[{Wei{\ss} {et~al.}(1999)Wei{\ss}, Walter, Neininger, \&
  Klein}]{weiss99}
Wei{\ss}, A., Walter, F., Neininger, N., \& Klein, U. 1999, \aap, 345, L23.
\newblock
  \url{https://ui.adsabs.harvard.edu/abs/1999A%26A...345L..23W/abstract}

\bibitem[{Williams {et~al.}(1994)Williams, {de Geus}, \& Blitz}]{williams94}
Williams, J.~P., {de Geus}, E.~J., \& Blitz, L. 1994, \apj, 428, 693,
  \dodoi{10.1086/174279}

\bibitem[{Willott {et~al.}(2015)Willott, Carilli, Wagg, \& Wang}]{willott15}
Willott, C.~J., Carilli, C.~L., Wagg, J., \& Wang, R. 2015, \apj, 807, 180,
  \dodoi{10.1088/0004-637X/807/2/180}

\bibitem[{Zavala {et~al.}(2018)Zavala, Monta{\~n}a, Hughes, Yun, Ivison,
  Valiante, Wilner, Spilker, Aretxaga, Eales, {Avila-Reese}, Ch{\'a}vez,
  Cooray, Dannerbauer, Dunlop, Dunne, {G{\'o}mez-Ruiz}, Micha{\l}owski,
  Narayanan, Nayyeri, Oteo, Rosa~Gonz{\'a}lez, {S{\'a}nchez-Arg{\"u}elles},
  Schloerb, Serjeant, Smith, Terlevich, Vega, Villalba, {van der Werf}, Wilson,
  \& Zeballos}]{zavala18}
Zavala, J.~A., Monta{\~n}a, A., Hughes, D.~H., {et~al.} 2018, Nature Astronomy,
  2, 56, \dodoi{10.1038/s41550-017-0297-8}

\end{thebibliography}

\end{CJK*}
\end{document}